\documentclass[reprint,aps,prl,twocolumns,superscriptaddress,floatfix,notitlepage]{revtex4-1}
\usepackage[a4paper]{geometry}
\usepackage{graphicx}
\usepackage{float}
\usepackage{amsmath}
\usepackage{amsfonts}
\usepackage{amssymb}
\usepackage{enumitem}
\usepackage{epstopdf}
\usepackage{mathrsfs}
\usepackage{mathtools}
\usepackage{fullpage}
\usepackage[utf8]{inputenc}
\usepackage{xcolor}
\usepackage{bbold}

\def\>{\rangle}
\def\<{\langle}

\newcommand\x{\mathbf{x}}

\newcommand\mk{\mathbf{k}}

\newcommand\mq{\mathbf{q}}

\newcommand{\be}{\begin{equation}}
\newcommand{\ee}{\end{equation}}
\newcommand{\bea}{\begin{eqnarray}}
\newcommand{\eea}{\end{eqnarray}}
\newcommand{\p}{\partial}
\newcommand{\1}{\frac{1}{2}}

\renewcommand{\vec}[1]{\mathbf{#1}}
\newcommand*\diff{\mathop{}\!\mathrm{d}}
\usepackage{braket}
\usepackage[colorlinks,bookmarks=true,citecolor=blue,linkcolor=red,urlcolor=blue]{hyperref}
\usepackage{hyperref}

\begin{document}
	\title{Transport in bilayer graphene near charge neutrality: Which scattering mechanisms are important?} 
	\author{Glenn Wagner}
	\affiliation{Rudolf Peierls Centre for Theoretical Physics, Parks Road, Oxford, OX1 3PU, UK}
	\author{Dung X. Nguyen}
	\affiliation{Rudolf Peierls Centre for Theoretical Physics, Parks Road, Oxford, OX1 3PU, UK}
	\author{Steven H. Simon}
	\affiliation{Rudolf Peierls Centre for Theoretical Physics, Parks Road, Oxford, OX1 3PU, UK}
	\begin{abstract} 
Using the semiclassical quantum Boltzmann equation (QBE), we numerically calculate the DC transport properties of bilayer graphene near charge neutrality.    We find, in contrast to prior discussions, that phonon scattering is crucial even at temperatures below 40K.  Nonetheless, electron-electron scattering still dominates over phonon collisions allowing a hydrodynamic approach.   We introduce a simple two-fluid hydrodynamic model of electrons and holes interacting via Coulomb drag and compare our results to the full QBE calculation.   We show that the two-fluid model produces quantitatively accurate results for conductivity, thermopower, and thermal conductivity.


	\end{abstract}
	
	\maketitle
    Two-dimensional materials such as graphene\cite{Novoselovaac9439} have attracted an enormous amount of attention in the last decade.  While the transport properties of graphene have been extensively studied\cite{NunoReview}, other related materials are only beginning to be examined.   Advances in nanotechnology have recently allowed electrical measurements on suspended bilayer graphene (BLG) samples\cite{Nam2017,sBLG1,sBLG2,sBLG3,sBLG4,sBLG5,sBLG6,sBLG7,sBLG8}. Motivated by these advances, we theoretically examine the DC transport properties of BLG using a quantum Boltzmann equation (QBE) approach.

	As in the case of monolayer graphene\cite{BOLOTIN2008351,Morozov2008}, we expect that electrons in BLG will have high mobility and scattering of electrons among themselves will be the dominant scattering mechanism.   We thus expect that the quantum Boltzmann equation (QBE) will reduce to some form of hydrodynamics\cite{Svintsov,Pellegrino2017,Hydro1,Hydro2,Hydro3,Hydro4,Mueller2008b}.


 
    Measurement of the electrical conductivity in BLG has been reported in Ref.~\cite{Nam2017} and we show that our QBE results agree with the experimental data over a wide range of parameters.       We focus our calculation on the regime explored in that experiment, i.e., close to charge neutrality and at temperatures $T\sim 10$K --  $40$K  (using the common Bernal stacked BLG with no out of plane field).    We show that in this regime, surprisingly, the effect of phonon scattering plays an essential role (with the effect of finite sample size playing a minor role).   We then use the QBE to make predictions for the thermopower and thermal conductivity that can verified in future experiments.    We then show that our transport results can be quantitatively understood with an extremely simple two-fluid hydrodynamic model\cite{Gantmakher1978,Levitov2013,Abanin2011,Titov2013}.  This simplification allows intuitive understanding of the physics that is not otherwise possible.

    

    
In this paper we begin by briefly discussing our QBE method, which we elaborate in detail in a companion paper\cite{Formalism2019}.  This approach is similar to previous work on the conductivity of monolayer graphene \cite{Fritz2008,Mueller2008}, two coupled monolayers \cite{Lux2012} and BLG\cite{Lux2013,Vignale}. Compared to \cite{Lux2013}, which studies the case of clean BLG, we calculate the conductivity away from CN, which requires including additional scattering mechanisms. The paper \cite{Vignale} does study the conductivity of BLG away from CN by including the effect of disorder, however we aim to provide a more quantitative analysis allowing for comparison with experimental data. We compare our results for the electrical conductivity to the experimental results from Ref.~\cite{Nam2017} in order to extract the value of the only free parameter of the theory: The phonon scattering rate. The value we determine for the phonon deformation potential lies within the range reported by previous authors. We then use this value to calculate the thermal conductivity and the thermo-electric coefficient. Finally, we introduce the two-fluid model and show that it agrees quantitatively with the more detailed QBE numerics. 
    
    \emph{QBE}---At low energies, bilayer graphene can be described in a two-band model with a quadratic dispersion $\epsilon_\lambda(k)=\lambda|\vec{k}|^2/2m$ where $\lambda=+(-)$ for electrons in the conduction (valence) band \cite{McCann2013,Formalism2019}, $m=0.033m_e$ is the effective mass \cite{Koshino2006}, and $\mk$ is the two-dimensional momentum.  There are two valleys  ($K$ and $K'$) and two spin states in each valley, thus giving $N_f = 4$ flavors. We neglect additional effects at low temperatures such as trigonal warping \cite{McCann2006} and opening of an interaction induced gap \cite{Nam2018}, restricting the applicability of our approach to $T\gtrsim10K$ \cite{Nam2017}. 
    
    The QBE is a semiclassical approach which is valid when the de Broglie wavelength is much shorter than the scattering lengths. This is equivalent to the condition $\alpha_x\equiv\beta\tau_x^{-1}\ll1$, where $\tau_x$ stands for the scattering time due to any of the scattering mechanisms in the problem, and $\beta$ is the inverse temperature.   
    
    In equilibrium the occupancy of a $\mk$ state is given by the usual Fermi factor.   The QBE determines the change in occupancy of the $\mk$ states due to small perturbations such as an external electric field $\vec{E}$ or temperature gradient $\nabla T$.  Formally the QBE is derived from the Kadanoff-Baym equations for the evolution of Green's functions using a Born collision integral to describe scattering.     Once Boltzmann equations are formally derived, the solution is obtained by expanding the $\bf k$ space occupation function $f({\bf k})$ in a set of basis functions, which then reduces the QBE to a set of linear equations which can be numerically solved. The size of the basis set is then expanded to convergence\cite{Dumitrescu:2015,Lux2013,Lux2012,Mueller2008}.  This method of solution for the Boltzmann equation is well-known in the plasma physics community and goes by the name Spitzer-H\"arm method \cite{Spitzer1953}.   Details of both the derivation of the QBE, and the method of solution are presented in detail in Ref.~\cite{Formalism2019} and the Supplementary  Material\cite{Supplement}.


	

We calculate the electrical current $\vec{J}$ and the heat current, which is defined as $\vec{Q}=\vec{J}^E-\frac{\mu}{e}\vec{J}$, where $\vec{J}^E$ is the energy current and $e<0$ is the electron charge. From this we can determine the electrical conductivity $\sigma$, the thermal conductivity $K$ and the thermo-electric coefficient $\Theta$ by
\begin{equation}
\begin{pmatrix}
\vec{J}\\
\vec{Q}
\end{pmatrix}
=\begin{pmatrix}
\sigma &\Theta\\
T\Theta& K
\end{pmatrix}
\begin{pmatrix}
\vec{E}\\
-\vec{\nabla}T
\end{pmatrix}_.
\label{eq:Def_coeff}
\end{equation}
The open circuit thermal conductivity \footnote{$\kappa$  measures the heat current in the absence of electrical current.} which is usually measured in experiments is given by $\kappa=K-T\Theta\sigma^{-1}\Theta$. 


    
\begin{figure}
\centering
\includegraphics[width=0.5\textwidth]{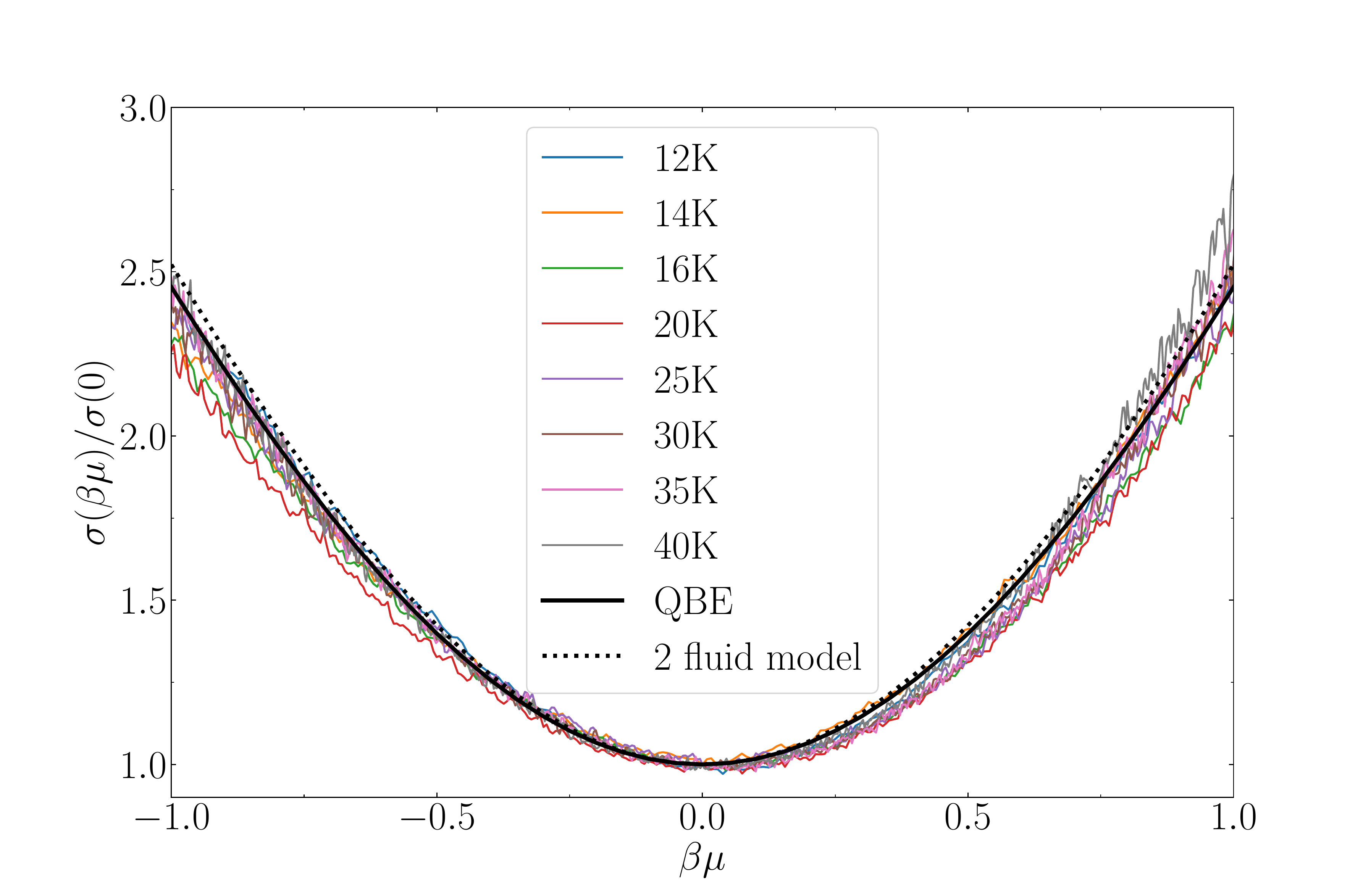}
\caption{Electrical conductivity $\sigma(\beta\mu)$ for different values of the temperature. From the experimental data, we fit the value $\alpha_{\textrm{ph}}=0.05$. The experimental data is from \cite{Nam2017}.  The black solid curve is the result of the QBE calculation. The black dashed curve shows the result from the two-fluid model and shows good agreement with the full QBE calculation.}
\label{fig:Sigma_xx}
\end{figure}

\emph{Coulomb and phonon scattering}---  We first consider Coulomb (electron-electron, electron-hole) scattering, which we expect to be dominant.    We use the form of the screened Coulomb potential valid in the experimentally-relevant regime $\beta\mu\lesssim 1$.  
In the experimental data\cite{Nam2017}, to a good approximation the conductivity only depends on the dimensionless combination $\beta\mu$.   (See Fig.~\ref{fig:Sigma_xx} \footnote{We also note that in \cite{Nam2017}, in three of four samples, $\sigma(\mu=0)$ does not vary more than about 10\% over the full range of measured temperatures $10-100$K.}).
If only Coulomb interactions are included then  $\beta\mu$ is  the only dimensionless parameter of the problem (the electromagnetic fine-structure constant cancels out when the screened potential is used\cite{Supplement}). 
However, away from charge neutrality (CN), the so-called {\it momentum mode} (i.e.,  a simple Galilean boost),  where electrons and holes move in the same direction, carries electrical current, and is not relaxed by Coulomb scattering since it conserves momentum. Thus, to obtain a finite conductivity, another scattering mechanism that relaxes momentum must be considered.  We identify three possible such mechanisms: Impurity scattering, scattering off the boundary of the finite-size sample, and phonons. In the former two cases, the relevant {\it dimensionless} scattering parameter $\alpha_x$ depends on temperature
\footnote{The dimensionless parameters for impurity scattering, scattering off the boundary and phonons respectively, are $\alpha_\textrm{imp}=(8\pi^2Z/N_f\epsilon_r)^2\beta n_{\textrm{imp}}/m$ $\alpha_L=\frac{\sqrt[]{\beta}}{\sqrt[]{m}L}$, $\alpha_{\textrm{ph}}=\beta \tau_\textrm{phonon}^{-1}$. Here $ L$ is the size of the sample, $Z$ is the charge of the impurities, $\epsilon_r$ is the relative permeability and $n_\textrm{imp}$ is the impurity density.}. Therefore, if either of these scattering mechanisms were most important (after Coulomb scattering) the curves for different temperatures would not collapse when plotted as a function of $\beta\mu$, and this does not agree with the experimental data.  We thus disregard these two scattering mechanisms. The situation, however, is different for phonon scattering.  The experiment is at temperatures above the Bloch-Gr\"uneisen temperature and hence the longitudinal acoustic phonons have a scattering time \cite{Viljas2010,Ochoa2011}
    \begin{equation}
        \tau_\textrm{ph}^{-1}=\frac{D^2m k_BT}{2\rho c^2},
    \end{equation}
where $c$ is the speed of sound in graphene, $D$ is the deformation potential and $\rho$ is the mass density of BLG. 
The relevant dimensionless parameter $\alpha_{\textrm{ph}}=\beta\tau_{\textrm{ph}}^{-1}$ is then independent of temperature and should result in the conductivity being a function of $\beta \mu$ only, in close agreement with experiment\cite{Nam2017}.      We emphasize the surprising result that even at these comparatively low temperatures of $12-40$K, phonons provide the primary momentum relaxation mechanism.

\emph{Electrical conductivity.---}   In the absence of a magnetic field the conductivity $\sigma_{ij}=\frac{N_fe^2}{2\hbar}\tilde\sigma\delta_{ij}$ depends on the dimensionless numbers $\beta\mu$ and $\alpha_\textrm{ph}$.    The thermal density of electrons (holes) for the free Fermi gas is given by $n^{e(h)} = (N_f m/(2 \pi \beta)) \ln(1 + e^{\pm \beta \mu})$ and as we increase $|\beta \mu|$ there are more charge carriers. However the total number density of electrons plus holes only increases by $17\%$  between $|\beta \mu|=0$ and 1 (at fixed $T$), and therefore this effect alone is too small to account for the electrical conductivity more than doubling in this range.    To understand this we realize that the primary scattering is electron-hole collisions\cite{Nam2017} and therefore the conductivity at large chemical potential is large both because we have many electrons to carry the charge and also because there are not many holes to scatter from.    A non-zero $\alpha_\textrm{ph}$ makes the conductivity finite and,  as mentioned above, the curves for different temperatures approximately collapse when plotted as a function of $\beta\mu$.  We treat the deformation potential $D$ as a fit parameter being that various different approaches have given different estimates of this quantity\cite{D1,D2,D3,Guinea2011,Borysenko2011}. Our best fit value $\alpha_\textrm{ph}=0.05$ corresponds to $D\approx 27$eV, which is consistent with prior expectation that it lies in the range $10-30$eV. Including the effect of finite size (boundary scattering) reduces the phonon scattering required to match the experiment. Taking account of the fact that the typical size of the system is around $3\mu m$, and there will be additional momentum-relaxing scattering off the boundary of the sample, the best fit $D$ may be reduced by around 30\%. 


In Fig.~\ref{fig:Sigma_xx} we show $\sigma$ as a function of $\beta\mu$.  We show both the result of QBE calculation using the above discussed fit value of $D$ as well as the experimental data from Ref.~\cite{Nam2017}.  

Exactly at CN, our prediction for the magnitude of $\sigma$ matches prior calculations by Ref.~\cite{Lux2013} to within about 1.5\% between 10K --100K.   Comparison to experiment is more difficult because different samples give different precise values of conductivity --- differing from each other by factors of up to about 4.   Our prediction lies acceptably in the middle of the experimental range. 

\begin{figure}
\centering
\includegraphics[width=0.5\textwidth]{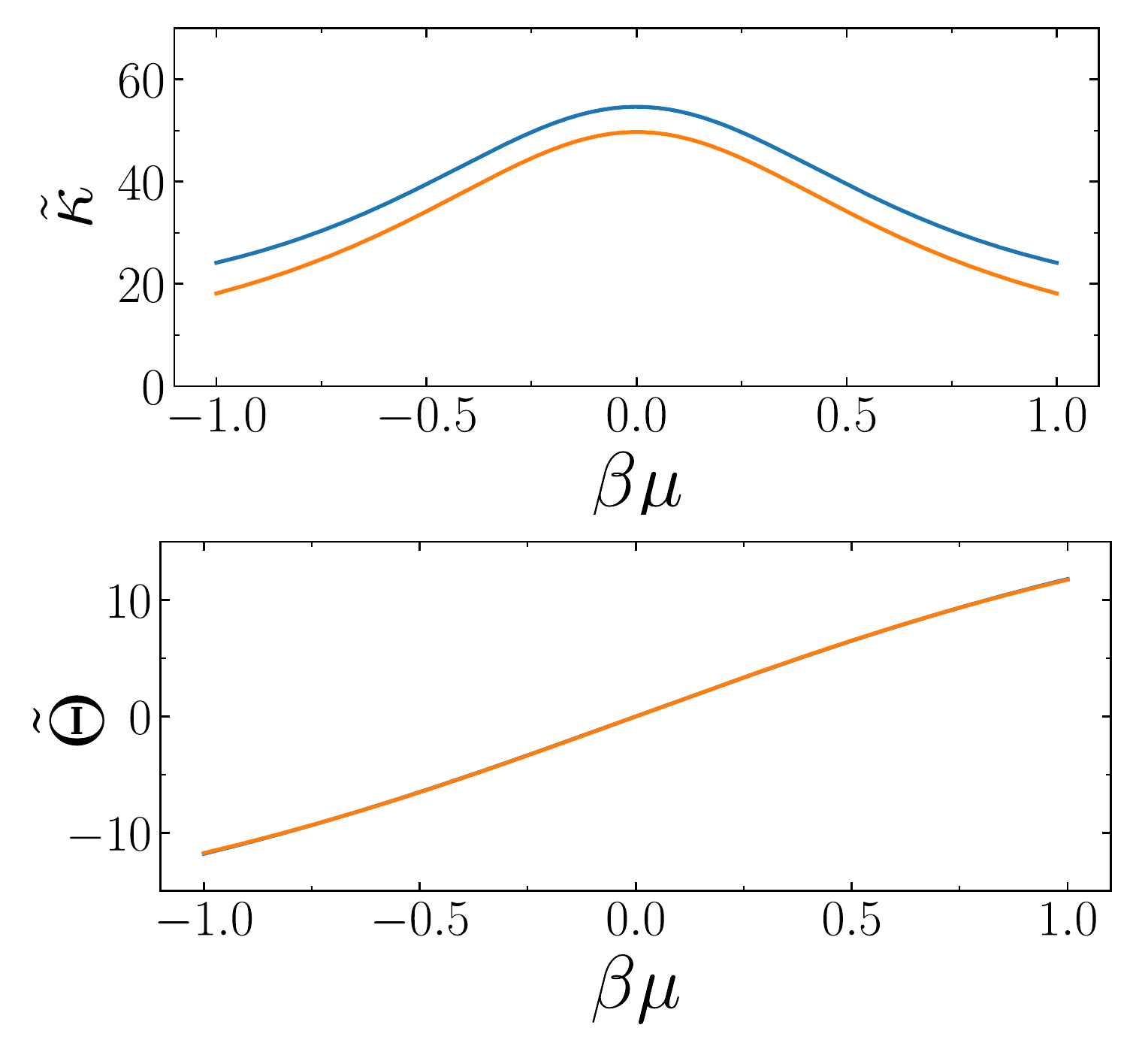}
\caption{Top: The dimensionless open-circuit thermal conductivity  $\tilde\kappa=(N_fk_B^2T/2\hbar)^{-1}\kappa$ calculated using QBE (blue) and two-fluid approximation (yellow).   Bottom: The dimensionless thermoelectric coefficient is defined as $\tilde\Theta=(N_fek_BT/2\hbar)^{-1}\Theta$ calculated using QBE (blue) and two-fluid approximation  (yellow).  The two curves show near perfect overlap.  Both figures use $\alpha_{\textrm{ph}}=0.05$. 
  }
\label{fig:Kappa_xx}
\label{fig:Theta_xx}
\end{figure}

\emph{Thermal Properties.}---Including only Coulomb scattering, the thermal conductivity $K$ diverges since the momentum mode carries thermal current and cannot be relaxed by Coulomb interactions.   Phonons again regulate this  divergence.  One might expect that the thermal conductivity increases with increasing $|\beta\mu|$ since the total heat transport carrier density $n^e+n^h$ increases. However, we note that $K$ actually decreases with increasing $|\beta \mu|$  which is counter-intuitive. In the limit of weak phonon scattering we will see below that $K \sim 1/(n^e+n^h)$. The open circuit thermal conductivity $\kappa_{ij}=\frac{N_fk_B^2T}{2\hbar} \tilde\kappa\delta_{ij}$, plotted in  Fig.~\ref{fig:Kappa_xx} (top), 
decreases faster than $K$ away from CN since the momentum mode carries electric current and hence does not contribute to $\kappa$.  The thermoelectric coefficient  $\Theta_{ij}=\frac{N_fek_BT}{2\hbar} \tilde\Theta\delta_{ij}$ plotted in Fig.~\ref{fig:Theta_xx} (bottom) vanishes at CN and increases as we increase $\beta\mu$, as the momentum mode now carries both heat and charge.

\emph{Two-fluid model.}---Since Coulomb scattering is dominant we expect a hydrodynamic description will be appropriate.   Further, since the scattering between electrons and holes is suppressed due to both matrix element effects and  energy-momentum conservation constraints\cite{Formalism2019}, we believe treating the electron fluid and the hole fluids as  weakly interacting with each other will be accurate.    In this limit, due to the strong scattering within each species, each of the two fluids should have a well defined temperature, chemical potential, and velocity.   We thus introduce a two-fluid model \cite{Levitov2013,Abanin2011,Titov2013,Gantmakher1978}, which shows excellent agreement with our detailed numerical calculation\footnote{We note that the two-fluid model is equivalent to performing the Spitzer-Harm calculation with two basis functions.}.    
The evolution of the mean fluid velocities of electrons (holes) $\vec{u}^e$ ($\vec{u}^h$) can then be derived explicitly from the QBE (See Supplementary Material\cite{Supplement} for detailed derivation) giving\footnote{By considering the Lorenz force we can also include the magnetic field in this equation.}.  
\begin{align} \nonumber
    m \dot{\bf u}^e &=-\frac{m}{\tau_{eh}}(\vec{u}^e-\vec{u}^h)- \frac{m}{\tau_{ph}}\vec{u}^e+ e\vec{E}-\Lambda^e k_B\nabla T \\ \nonumber
    m  \dot{\bf u}^h &=-\frac{m}{\tau_{he}}(\vec{u}^h-\vec{u}^e)- \frac{m}{\tau_{ph}}\vec{u}^h- e\vec{E}-\Lambda^h k_B\nabla T
\end{align}
 Here $\tau_{eh} = \tau_0 (n^e + n^h)/n^h$ is the scattering time for electrons to scatter off holes and $n^e \tau_{eh} = n^h \tau_{he}$ guarantees momentum conservation.  By evaluating the Coulomb collision integral we extract the value $\beta \tau_0^{-1} = 0.15$ (See Supplement\cite{Supplement}).       Here, $\Lambda^{e}\ (\Lambda^{h})$ is the entropy per electron (hole) of  the free Fermi gas characterized by $\beta \mu$. An explicit expression for these quantities is given in Supplementary Material\cite{Supplement}. 

We are interested in DC transport so we may set the left of these two equations to zero, and solve for the velocities ${\bf u}^{e,h}$.  The electric current is then ${\bf J} =  e(n^e {\bf u}^e - n^h {\bf u}^h)$ and the thermal current is ${\bf J}^E = k_B T (\Lambda^e n^e {\bf u}^e + \Lambda^h n^h {\bf u}^h)$.   The advantage of the two-fluid model is its simplicity which allows us to obtain analytic expressions (see Supplement\cite{Supplement}) for transport coefficients.  In the limit $\tau_{ph} \gg \tau_0$ the expressions simplify even more giving, for example, $\tilde K = (2 \pi^2 m k_B T/9) \alpha_{ph}^{-1} / (n^e + n^h)$.     In Figs.~\ref{fig:Sigma_xx} and \ref{fig:Kappa_xx} we compare the results from the QBE and the two-fluid model.  In both cases the agreement is extremely good.



Since heat is carried by the momentum mode, the Coulomb drag between electrons and holes will not affect it, and the thermal conductivity will mainly be limited by the phonon scattering. The electrical conductivity on the other hand will be limited by both  phonon scattering and the Coulomb drag between electrons and holes. Therefore, the electrical conductivity is suppressed compared to the thermal conductivity, leading to a large Lorenz number $\mathcal{L}\equiv \kappa/(\sigma T)$. At charge neutrality we find $\mathcal{L} \approx 25(k_B/e)^2$ which is much larger than the value of $\pi^2/3(k_B/e)^2$ predicted by the Wiedemann-Franz law. The violation of the Wiedemann-Franz law for BLG has been previously pointed out by \cite{Vignale} and has been experimentally observed for monolayer graphene in \cite{WFviolation}.

\emph{Conclusion.}--- In this paper we have calculated the transport properties of bilayer graphene. Our results for the electrical conductivity match the experimental results in \citep{Nam2017}. From the experimental data we deduce that even at low temperatures, the scattering off phonons is crucial.  Nonetheless, the dominant scattering mechanism is between charge carriers of the same species, which justifies a two fluid approach which shows excellent agreement with the detailed numerical results of the QBE and provides a simple way of calculating the transport properties analytically. One can adapt our two fluid model to different experimental setups with slight modifications \cite{Formalism2019}.


It would be interesting to test the predictions for the thermopower and thermal conductivity in upcoming experiments. 
Our formalism can be expanded to address a variety of other quantities of interest, including finite frequency effects, spin transport, and Hall viscosity.

\emph{Acknowledgements.}--- We would like to acknowledge helpful discussions with Philipp Dumitrescu and Lars Fritz. We would also like to thank the authors of \cite{Nam2017} for kindly providing the data from that work to us. We are very grateful to David Soler Delgado and Alberto Morpurgo for discussing their experiment with us. This work was supported by EP/N01930X/1. Statement  of  compliance  with  EPSRC  policy  framework  on  research  data:  This  publication is theoretical work that does not require supporting research data.

%

	\newpage

	\begin{widetext}
	\begin{center}
		\textbf{\large --- Supplementary Material ---\\ Transport in bilayer graphene near charge neutrality: Which scattering mechanisms are important?}\\
		\medskip
		\text{Glenn Wagner, Dung X. Nguyen and Steven Simon }
	\end{center}
	\setcounter{equation}{0}
	\setcounter{figure}{0}
	\setcounter{table}{0}
	\setcounter{page}{1}
	\makeatletter
	\renewcommand{\theequation}{S\arabic{equation}}
	\renewcommand{\thefigure}{S\arabic{figure}}
	\renewcommand{\bibnumfmt}[1]{[S#1]}

\section{Summary of the Boltzmann equation formalism}
In this section, we briefly summarize the formalism \cite{Formalism2019} used to calculate transport coefficients in BLG near the neutral point. To calculate the transport coefficients, we solve the QBE equation in the perturbed background to obtain the perturbation of distribution function $f_\lambda(\mathbf{p})$ and then use the result to derive the linear response. The Boltzmann equation is
	\begin{equation}
	\label{eq:Boltz1}
	\left(\frac{\p}{\p t}+\mathbf{v}_\lambda(\mathbf{k}) \cdot\frac{\p}{\p \x} +e\mathbf{E}(\mathbf{x},t)  \cdot \frac{\p}{\p \mk}\right)f_{\lambda}(\mk,\x,t)\\=-I_{\lambda}[\{f_{\lambda_i}\}](\mk,\mathbf{x},t),
	\end{equation}
	where the group velocity of band $\lambda$ is defined as 
	\begin{equation}
	\mathbf{v}_{\lambda}(\mathbf{k})=\p_{\mathbf{k}}\epsilon_{\lambda}(\mathbf{k}).
	\label{eq:v_group}
	\end{equation}
The distribution function is expanded as
	\begin{equation}
	f_\lambda(\mathbf{p})=f^0_\lambda(\mathbf{p})+f^0_\lambda(\mk)[1-f^0_\lambda(\mathbf{p}))]h_\lambda(\mathbf{p}).
	\label{eq:f_pert}
	\end{equation}	
about the equilibrium Fermi distribution 
	\begin{equation}
	f^0_\lambda(\mathbf{p})=\frac{1}{1+e^{\beta(\epsilon_\lambda(\mathbf{p})-\mu)}},
	\label{eq:f_0}
	\end{equation}
where $\epsilon_\lambda(\mathbf{p})=\lambda p^2/(2m)$ is the energy in band $\lambda$. The linearized quantum Boltzmann equation can then be derived to be
\begin{equation}
-I_\lambda[\{h_{\lambda_i}(\vec k_i)\}](\mathbf{p})=-\frac{\lambda\beta}{m}f^0_\lambda(\mathbf{p})[1-f^0_\lambda(\mathbf{p})]\bigg(e\mathbf{E}\cdot\mathbf{p}-\frac{1}{T}\nabla T\cdot\mathbf{p}(\epsilon_\lambda(\mathbf{p})-\mu)\bigg),
\label{eq:QBE_app}
\end{equation} 
where $\vec E$ is the electric field and $\nabla T$ is the thermal gradient.
The collision integral includes contributions from quasi-particle interactions, scattering on phonons  and finite size effect respectively
\begin{equation}
	\label{eq:CoItot}
	I_{\lambda}[h_{\lambda_i} (\vec{k}_i)](\mathbf{p})=I_{\lambda,\textrm{int}}[h_{\lambda_i} (\vec{k}_i)](\mathbf{p})+I_{\lambda,\textrm{phonon}}[h_{\lambda_i} (\vec{k}_i)](\mathbf{p})+I_{\lambda,\textrm{size}}[h_{\lambda_i} (\vec{k}_i)](\mathbf{p}).
\end{equation}
The first term in \eqref{eq:CoItot} accounts for the Coulomb scattering of charge carriers (electrons and holes) off each other. The second term describes the scattering of charge carriers off acoustic phonons. The last term accounts for scattering of the charge carriers off the boundary of the necessarily finite-size sample. The linearized collision integral for scattering between charge carriers is derived in \cite{Formalism2019} as
\begin{align}
\label{eq:CoI1}
I_{\lambda,\textrm{int}}^{(1)}[h_{\lambda_i} (\vec{k}_i)](\vec{p})&=-(2\pi)\sum_{\lambda_1\lambda_2\lambda_3}\int \frac{\diff^2\vec{k}}{(2\pi)^2} \frac{\diff^2\vec{q}}{(2\pi)^2}\delta(\epsilon_{\lambda}(\vec{p})+\epsilon_{\lambda_1}(\vec{k})-\epsilon_{\lambda_2}(\vec{p+q})-\epsilon_{\lambda_3}(\vec{k-q}))\nonumber\\
&\times\bigg[N_f|T_{\lambda\lambda_1\lambda_3\lambda_2}(\vec{p},\vec{k},\vec{q})|^2-T_{\lambda\lambda_1\lambda_3\lambda_2}(\vec{p},\vec{k},\vec{q})T_{\lambda\lambda_1\lambda_2\lambda_3}^*(\vec{p},\vec{k},\vec{k-p-q})\bigg]\nonumber \\
&\times \bigg[[1-f_\lambda^0(\vec{p})][1-f_{\lambda_1}^0(\vec{k})]f_{\lambda_2}^0(\vec{p+q})f_{\lambda_3}^0(\vec{k-q})\bigg]\nonumber\\
&\times\bigg[-h_{\lambda}(\vec{p})-h_{\lambda_1}(\vec{k})+h_{\lambda_2}(\vec{p+q})+h_{\lambda_3}(\vec{k-q})\bigg].
\end{align}
$N_f=4$ is the number of fermion species and the matrix elements are
\begin{equation}
	T_{\lambda_1\lambda_2\lambda_3\lambda_4}(\mk,\mk',\mq)=V_C(-\mq)M_{\lambda_1\lambda_4}(\mk+\mq,\mk)M_{\lambda_2\lambda_3}(\mk'-\mq,\mk')
\end{equation}
with
	\begin{equation}
	M_{\lambda\lambda'}(\mk,\mk')=\frac{1}{2}\left(1+\lambda\lambda'e^{i(2\theta_{\mk'}-2\theta_{\mk})}\right)
	\end{equation}
where $\theta_\mk$ is the angle between $\mk$ and the x-axis. The screened Coulomb potential is given according to \cite{Formalism2019} by 
\begin{equation}
V_C(q)=\frac{2\pi \alpha_{EM}}{q+q_{TF}(q)}
\label{eq:screened_V}
\end{equation}
with the Thomas-Fermi screening momentum 
\begin{equation}
\label{eq:qTF}
q_{TF}(q)= \alpha_{EM} m N_f(1+\frac{\beta q^2}{12m}),
\end{equation}
where $\alpha_{EM}=e^2/\epsilon_0$ is the electromagnetic fine structure constant.  In Eq.~\ref{eq:screened_V} we can take the denominator to be only $q_{TF}$ if $q \ll \alpha_{EM} m N_F$.  This condition is equivalent to the de Broglie wavelength being much greater than the Bohr radius ($\sim$ 15 Angstrom), which is true for the entire temperature range we consider.   We can thus take the approximation $V_C(q)=2\pi/N_fm(1+\frac{\beta q^2}{12m})^{-1}$. Note that the electromagnetic fine structure constant has dropped out when calculating the screened potential. The contribution to collision integral due to scattering off phonons is
\begin{equation}
	I_{\lambda,\textrm{phonon}}[h_{\lambda_i} (\vec{k}_i)](\vec{p})=\tau_\textrm{phonon}^{-1}f_\lambda^0(p)[1-f_\lambda^0(p)]h_\lambda(\vec{p})
	\end{equation}
and due to the finite size effect is
	\begin{equation}
	\label{eq:CoS}
	I_{\lambda,\textrm{size}}[h_{\lambda_i} (\vec{k}_i)](\vec{p})=\frac{p}{ mL}f_\lambda^0(p)[1-f_\lambda^0(p)]h_\lambda(\vec{p}),
	\end{equation}
where $L$ is a characteristic length scale of the sample. We present more detailed numerical calculations for each transport coefficient in the companion paper \cite{Formalism2019}. We can define the dimensionless parameters $\alpha_L=\frac{\sqrt[]{\beta}}{\sqrt[]{m}L}$ and $\alpha_{\textrm{ph}}=\beta \tau_\textrm{phonon}^{-1}$ to characterize the strength of the two scattering mechanisms. Here, $\alpha_x$ is the ratio of the de Broglie wavelength of a particle travelling at the thermal speed divided by the mean free path of the scattering mechanism $x$.

\section{Two-fluid model}
We introduce the two-fluid model, which reproduces the salient features of our numerical results. We start with the Boltzmann equation \eqref{eq:QBE_app}. We multiply the Boltzmann equation by $\lambda \vec{p}/m$ and integrate over momentum space in order to derive the evolution of the mean fluid velocities as

\begin{equation}
    m\partial_t\vec{u}^e=-\frac{m}{\tau_{eh}}(\vec{u}^e-\vec{u}^h)-\frac{m\vec{u}^e}{\tau_{ph}}-\frac{m\vec{u}^e}{\tau_{Le}}+e\vec{E}-\Lambda^e k_B\nabla T
    \label{eq:fluid1}
\end{equation}

\begin{equation}
    m\partial_t\vec{u}^h=\frac{m}{\tau_{he}}(\vec{u}^e-\vec{u}^h)-\frac{m\vec{u}^h}{\tau_{ph}}-\frac{m\vec{u}^h}{\tau_{Lh}}-e\vec{E}-\Lambda^h k_B\nabla T,
        \label{eq:fluid2}
\end{equation}
where $\tau_{ph}$ is the scattering time due to collisions with phonons. $\tau_{eh}$ and $\tau_{Le}$ are the scattering times of electrons by holes and by the boundary respectively. $\tau_{he}$ and $\tau_{Lh}$ are the scattering times of holes by electrons and by the boundary respectively. We defined the electron and hole velocities as 
\begin{equation}
\mathbf{u}^e=\frac{\int \frac{d^2 \vec{p}}{(2\pi)^2}\frac{\mathbf{p}}{m} f_+(\mathbf{p})}{\int \frac{d^2 \vec{p}}{(2\pi)^2} f^0_+(\mathbf{p})},\qquad \mathbf{u}^h=-\frac{\int \frac{d^2 \vec{p}}{(2\pi)^2}\frac{\mathbf{p}}{m} (1-f_-(\mathbf{p}))}{\int \frac{d^2 \vec{p}}{(2\pi)^2} (1-f^0_-(\mathbf{p}))}.
\end{equation}
The scattering times off the boundary are
\begin{equation}
\label{eq:tauLE}
    \tau_{Le}^{-1}=\frac{\beta}{2m^2L}\frac{\int \frac{d^2 \vec{p}}{(2\pi)^2} p^3 f^0_+(\mathbf{p})[1-f^0_+(\mathbf{p})]}{\int \frac{d^2 \vec{p}}{(2\pi)^2} f^0_+(\mathbf{p})}
\end{equation}
\begin{equation}
\label{eq:tauLH}
    \tau_{Lh}^{-1}=\frac{\beta}{2m^2L}\frac{\int \frac{d^2 \vec{p}}{(2\pi)^2} p^3 f^0_-(\mathbf{p})[1-f^0_-(\mathbf{p})]}{\int \frac{d^2 \vec{p}}{(2\pi)^2} (1-f^0_-(\mathbf{p}))}
\end{equation}
The coefficients $\Lambda^{e,h}$ account for the fact that the average entropy per particle is $\Lambda k_B$ 

\begin{equation}
\label{eq:Lambdae}
    k_BT\Lambda^e=\frac{\int \frac{d^2 \vec{p}}{(2\pi)^2} p^2(\epsilon_+(p)-\mu)f^0_+(\mathbf{p})[1-f^0_+(\mathbf{p})]}{\int \frac{d^2 \vec{p}}{(2\pi)^2} f^0_+(\mathbf{p})}
\end{equation}

\begin{equation}
\label{eq:Lambdah}
    k_BT\Lambda^h=\frac{\int \frac{d^2 \vec{p}}{(2\pi)^2} p^2(-\epsilon_-(p)+\mu)f^0_-(\mathbf{p})[1-f^0_-(\mathbf{p})]}{\int \frac{d^2 \vec{p}}{(2\pi)^2}(1-f^0_-(\mathbf{p}))}
\end{equation}
One can derive explicitly equations \eqref{eq:Lambdae} and \eqref{eq:Lambdah} from the $\nabla T$ term in Boltzmann's equation \eqref{eq:QBE_app}. It is easy to show that these integrals are in fact the entropy per particle 
\begin{equation} 
\Lambda^e=-\frac{\int \frac{d^{2}\vec p}{(2 \pi)^{2}}\left[\left(1-f^0_+(\mathbf{p})\right) \ln \left(1-f^0_+(\mathbf{p})\right)+f^0_+(\mathbf{p}) \ln f^0_+(\mathbf{p}) \right]}{\int \frac{d^{2} \vec p}{(2 \pi)^{2}} f^0_+(\mathbf{p})} ,
\end{equation}

\begin{equation} 
\Lambda^h=-\frac{\int \frac{d^{2}\vec p}{(2 \pi)^{2}}\left[\left(1-f^0_-(\mathbf{p})\right) \ln \left(1-f^0_-(\mathbf{p})\right)+f^0_-(\mathbf{p}) \ln f^0_-(\mathbf{p}) \right]}{\int \frac{d^{2} \vec p}{(2 \pi)^{2}} [1-f^0_-(\mathbf{p})]} ,
\end{equation}
which were used previously in \cite{Levitov2013}. The Coulomb drag term can be derived explicitly from the collision integral
\begin{equation}
\int \frac{d^2 \vec{p}}{(2\pi)^2} \frac{\lambda\vec{p}}{m}I_{\lambda,\textrm{int}}^{(1)}\bigg[h_{\lambda_i} (\vec{k}_i)=\lambda_i\beta \vec{k}_i\cdot \bigg(\frac{\vec{u}^e-\vec{u}^h}{2}\bigg)\bigg](\vec{p})=
\left\{\begin{array}{ll}-\frac{mn^e}{\tau_{eh}}(\vec{u}^e-\vec{u}^h)   & \lambda=+ \\ 
\frac{mn^h}{\tau_{he}}(\vec{u}^e-\vec{u}^h) & \lambda=-
\end{array} \right. 
\label{eq:Coul_drag}
\end{equation} 
This allows us to calculate $\tau_{eh}$ and $\tau_{he}$. We perform the calculation at charge neutrality and then use \eqref{eq:away_from_CN} to extrapolate. $\tau_{ph}$ is taken from the fit with the experimental data (the deformation potential $D$ being fit to experiment, as discussed in the main text). Aside from this, our model has no free parameters. We consider the steady state $\partial_t\vec{u}^e=\partial_t\vec{u}^h=0$ and solve the fluid equations \eqref{eq:fluid1} and \eqref{eq:fluid2} for $\vec{u}^e$ and $\vec{u}^h$. Using these solutions, we calculate the electric current 

\begin{equation}
    \vec{J}=e(n^e\vec{u}^e-n^h\vec{u}^h)
\end{equation}
where $e<0$ is the electron charge. The number densities calculated from the Fermi distribution are

\begin{equation}
    n^e=\frac{N_fm}{2\pi\beta}\ln(1+e^{\beta\mu}), \qquad n^h=\frac{N_fm}{2\pi\beta}\ln(1+e^{-\beta\mu})
\end{equation}
The heat current is defined as $\vec{Q}=\vec{J}^E-\frac{\mu}{e}\vec{J}$, where $\vec{J}^E$ is the energy current. Thus, the heat current is given by
\begin{equation}
    \vec{Q}= k_BT(\Lambda^en^e\vec{u}^e+\Lambda^hn^h\vec{u}^h)
\end{equation}
We define the electrical conductivity $\sigma$, the thermal conductivity $K$ and the thermo-electric coefficient $\Theta$ by
\begin{equation}
\begin{pmatrix}
\vec{J}\\
\vec{Q}
\end{pmatrix}
=\begin{pmatrix}
\sigma &\Theta\\
T\Theta& K
\end{pmatrix}
\begin{pmatrix}
\vec{E}\\
-\vec{\nabla}T
\end{pmatrix}_.
\label{eq:Def_coeff2}
\end{equation}
The open circuit thermal conductivity $\kappa$ measuring the heat current in the absence of electrical current and which is usually measured in experiments is given by $\kappa=K-T\Theta\sigma^{-1}\Theta$.  With these definitions we can derive  the transport coefficients from the two fluid model:
\begin{equation}
    \sigma=\frac{e^2\bigg(n^e+n^h+\tau_{ph}[n^e\tau_{Lh}^{-1}+n^h\tau_{Le}^{-1}+(\tau_{he}^{-1}-\tau_{eh}^{-1})(n^e-n^h)]\bigg)}{m(\tau_{ph}^{-1}+\tau_{eh}^{-1}+\tau_{he}^{-1}+\tau_{Lh}^{-1}+\tau_{Le}^{-1}+\tau_{ph}(\tau_{eh}^{-1}\tau_{Lh}^{-1}+\tau_{he}^{-1}\tau_{Le}^{-1}+\tau_{Le}^{-1}\tau_{Lh}^{-1}))}
\end{equation}

\begin{equation}
    \Theta=\frac{ek_B\bigg(n^e\tilde\Lambda^e-n^h\tilde\Lambda^h\bigg)}{m(\tau_{ph}^{-1}+\tau_{eh}^{-1}+\tau_{he}^{-1}+\tau_{Lh}^{-1}+\tau_{Le}^{-1}+\tau_{ph}(\tau_{eh}^{-1}\tau_{Lh}^{-1}+\tau_{he}^{-1}\tau_{Le}^{-1}+\tau_{Le}^{-1}\tau_{Lh}^{-1}))}
\end{equation}

\begin{equation}
    K=\frac{k_B^2T\bigg(\Lambda^en^e\tilde\Lambda^e+\Lambda^hn^h\tilde\Lambda^h\bigg)}{m(\tau_{ph}^{-1}+\tau_{eh}^{-1}+\tau_{he}^{-1}+\tau_{Lh}^{-1}+\tau_{Le}^{-1}+\tau_{ph}(\tau_{eh}^{-1}\tau_{Lh}^{-1}+\tau_{he}^{-1}\tau_{Le}^{-1}+\tau_{Le}^{-1}\tau_{Lh}^{-1}))}
\end{equation}
where
\begin{equation}
    \tilde\Lambda^e=\Lambda^e(1+\tau_{ph}(\tau_{he}^{-1}+\tau_{Lh}^{-1}))+\Lambda^h\tau_{ph}\tau_{eh}^{-1}
\end{equation}

\begin{equation}
    \tilde\Lambda^h=\Lambda^h(1+\tau_{ph}(\tau_{eh}^{-1}+\tau_{Le}^{-1}))+\Lambda^e\tau_{ph}\tau_{he}^{-1}
\end{equation}
Neglecting the scattering off the boundary we find the simpler expressions
\begin{equation}
    \sigma=\frac{e^2}{m(\tau_{ph}^{-1}+\tau_{eh}^{-1}+\tau_{he}^{-1})}\bigg(n^e+n^h+\tau_{ph}(\tau_{he}^{-1}-\tau_{eh}^{-1})(n^e-n^h)\bigg)
\end{equation}

\begin{equation}
    \Theta=\frac{ek_B}{m(\tau_{ph}^{-1}+\tau_{eh}^{-1}+\tau_{he}^{-1})}\bigg(n^e\tilde\Lambda^e-n^h\tilde\Lambda^h\bigg)
\end{equation}

\begin{equation}
    K=\frac{k_B^2T}{m(\tau_{ph}^{-1}+\tau_{eh}^{-1}+\tau_{he}^{-1})}\bigg(\Lambda^en^e\tilde\Lambda^e+\Lambda^hn^h\tilde\Lambda^h\bigg)
\end{equation}
where
\begin{equation}
    \tilde\Lambda^e=\Lambda^e(1+\tau_{ph}\tau_{he}^{-1})+\Lambda^h\tau_{ph}\tau_{eh}^{-1}
\end{equation}

\begin{equation}
    \tilde\Lambda^h=\Lambda^h(1+\tau_{ph}\tau_{eh}^{-1})+\Lambda^e\tau_{ph}\tau_{he}^{-1}
\end{equation}
For momentum conservation we require
\begin{equation}
    n^e\tau_{eh}=n^h\tau_{he}
    \label{eq:mmtm_cons}
\end{equation}
We verify explicitly that the Onsager relations for the thermoelectric coefficients are satisfied if equation \eqref{eq:mmtm_cons} is satisfied.
Thus we can choose
\begin{equation}
    \tau_{eh}=\tau_0\frac{n^e+n^h}{n^h}, \qquad \tau_{he}=\tau_0\frac{n^e+n^h}{n^e}
    \label{eq:away_from_CN}
\end{equation} 
By evaluating the collision integral in \eqref{eq:Coul_drag} at CN numerically, we find

\begin{equation}
    \alpha_0\equiv \beta\tau_0^{-1}=0.15
\end{equation}
From the experimental data (see main text) we extracted $\alpha_{\textrm{ph}}=0.05$ and from $L\sim 3\mu$m we have $\alpha_{L}=0.03/\sqrt{T[K]}<0.01$ for $T>10K$. Therefore we find that $\alpha_{\textrm{ph}},\alpha_{\textrm{L}}\ll \alpha_0$ which means that the scattering rate due to electron-electron collisions dominates over the electron-phonon scattering and the scattering of electrons off the boundary. This justifies the use of the hydrodynamic theory.
In the limit $\tau_\textrm{ph}\gg\tau_0$ and again neglecting the scattering off the boundary we find the simpler expressions

\begin{equation}
    \tilde\sigma\equiv \frac{2\hbar}{N_f e^2}\sigma=\frac{\beta}{2m}\bigg(\alpha_0^{-1}(n^e+n^h)+\alpha_{ph}^{-1}\frac{(n^e-n^h)^2}{n^e+n^h}\bigg)
\end{equation}

\begin{equation}
    \tilde\Theta\equiv \frac{2\hbar}{N_f ek_BT}\Theta=\frac{\pi}{3}\alpha_{ph}^{-1}\frac{n^e-n^h}{n^e+n^h}
\end{equation}

\begin{equation}
    \tilde K\equiv \frac{2\hbar}{N_f k_B^2T}K=\frac{2\pi^2}{9}\alpha_{ph}^{-1}\frac{m}{\beta(n^e+n^h)}
\end{equation}
where the tilde quantities are dimensionless as defined in the main text. 
In deriving the above equation we have made use of the identity
\begin{equation}
    n^e\Lambda^e+n^h\Lambda^h=\frac{2\pi}{3}\frac{m}{\beta},
\end{equation}
which can be proven from the explicit integral expressions for $\Lambda^{e,h}$.

\section{Supplementary figures}

\begin{figure}[H]
\centering
\includegraphics[width=0.5\textwidth]{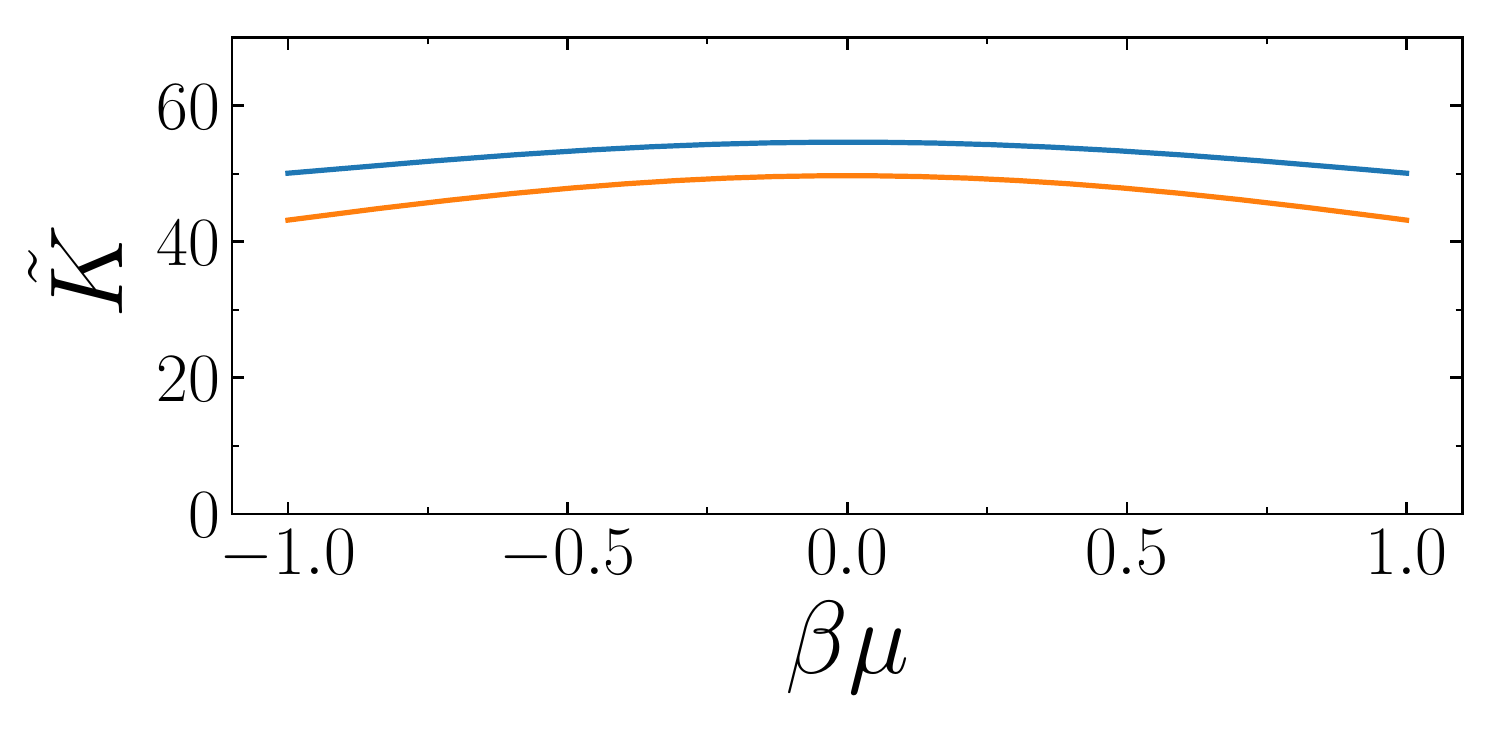}
\caption{Thermal conductivity coefficient $\tilde K(\beta\mu)$ (blue) and two-fluid approximation to $\tilde{K}(\beta\mu)$ (yellow) for $\alpha_{\textrm{ph}}=0.05$. We have defined $K_{ij}=\frac{N_fk_B^2T}{2\hbar} \tilde K\delta_{ij}$. The prediction from the two-fluid model and the detailed numerical calculation roughly differ by a constant.}
\end{figure}




\end{widetext}
\end{document}